\documentclass[prl,groupedaddress,preprint,draft,onecolumn,12pt,notitlepage,tightenlines]{revtex4}
\input{tcilatex}

\begin{document}

\title{Comment on ``A linear optics implementation of weak values in Hardy's
paradox''}
\author{J.S. Lundeen$^{1}$, K.J. Resch$^{2}$ and A.M. Steinberg$^{1,2}$}
\affiliation{$^{1}$Department of Physics, University of Toronto, 60 St. George Street,
Toronto ON M5S 1A7, CANADA\\
$^{2}$Institut f\"{u}r Experimentalphysik, Universit\"{a}t Wien \\
Boltzmanngasse 5, A-1090 Vienna, AUSTRIA}

\begin{abstract}
A recent experimental proposal by Ahnert and Payne\emph{\ }[S.E. Ahnert and
M.C. Payne, Phys. Rev. A \textbf{70}, 042102 (2004)] outlines a method to
measure the weak value predictions of Aharonov in Hardy's paradox. \ This
proposal contains flaws such as the state preparation method and the
procedure for carrying out the requisite weak measurements.\emph{\ \ }We
identify previously published solutions to some of the flaws.
\end{abstract}

\maketitle

\section{Introduction}

Weak measurement is an extension of the von Neumann measurement model in
which the coupling of the measurement pointer to the measured system is
small \cite{Aharanov1}. \ This prevents measurement-induced disturbance of
the measured system and avoids collapse. \ Consequently, weak measurement is
suitable for studying systems involving post-selection such as Hardy's
paradox \cite{Hardy}. \ Hardy's paradox involves two overlapped Mach-Zehnder
interferometers where each interferometer acts as an interaction-free
measurement (IFM) \cite{IFM} on the particle in the ``inner'' arm of other
interferometer. \ The paradoxical result is that occasionally both
interaction-free measurements are positive and indicate the simultaneous
presence of the two photons in the inner arms of the Mach-Zehnders yet one
never finds the photon-pair there when directly measured. \ Aharonov used
weak measurement to find which Mach-Zehnder arms the photons were in,
individually and as a pair, in the subensemble of systems for which the IFMs
give their paradoxical result \cite{Aharanov}. \ Unlike standard strong
quantum measurements, the results of these weak measurements, weak values,
are consistent with each other and hence resolve the paradox of where the
photon pair was. \ The resolution is nevertheless strange as it shows that
there are negative one pairs of photons in the ``outer'' arms. \ Although
there has been a proposal to test Aharonov's predictions in an ion system %
\cite{Molmer}, as of yet there have been no experiments.\ \ This comment is
on a recent experimental proposal by Ahnert and Payne for a linear optics
implementation of Hardy's paradox and the aforementioned weak measurements %
\cite{Ahnert}. \ We address two main problems in the proposal. \ First, the
post-selected linear optics state preparation procedure does not produce the
correct state for Hardy's paradox. \ And second, the outlined methods for
conducting weak measurements of two-particle observables are not capable of
measuring the weak values of the operators Aharonov investigated.

\section{State Preparation Procedure}

In this section, we show that Ahnert and Payne's procedure for creating the
necessary state with linear optics and post-selection does not work. \ The
authors choose a polarization representation to encode the paths of the
photons, where $\left| H\right\rangle $ represents the Mach-Zehnder inner
path and $\left| V\right\rangle $\ represents the outer. \ They aim to
produce the nonmaximally-entangled initial state $\left( \left|
HH\right\rangle +\left| HV\right\rangle +\left| VH\right\rangle \right) /%
\sqrt{3}$.\ \ They begin with two pairs of photons, each in the state $%
\left( \left| HH\right\rangle +\left| VV\right\rangle \right) /\sqrt{2}$,
that enter the setup shown in Fig. 3 of Ref. \cite{Ahnert}. \ Upon a
detection \ at $D^{\prime }$, the authors claim the state collapses to the
target initial state. \ In fact, their apparatus produces the density matrix 
$\left( \left| HH\right\rangle \left\langle HH\right| +\left|
HV\right\rangle \left\langle HV\right| +\left| VH\right\rangle \left\langle
VH\right| \right) /3.$ \ The reason that there is no coherence between the
terms in this state is because the apparatus cannot remove the which-path
information remaining after the detection of a photon at $D^{\prime }$: \ $%
\left| HH\right\rangle $ results in no photons exiting the two polarizing
beam splitters (PBS); $\left| HV\right\rangle $ results in a V photon
exiting the upper PBS; and $\left| VH\right\rangle $ results in V photon
exiting the lower PBS. \ Tracing over the modes exiting the vertical ports
of the two PBSs leaves us with a mixed state unsuitable for Hardy's paradox.
\ In fact, this flaw is a relatively minor problem as there are other even
simpler methods that require only one pair of photons and linear optics to
produce the appropriate entangled state. \ Specifically, the target initial
state can be written as $a\left| \psi \phi \right\rangle +b\left| \psi
^{\perp }\phi ^{\perp }\right\rangle $ via a Schmidt decomposition.\ \
Beginning with a source that emits a photon pair in $a\left| HH\right\rangle
+b\left| VV\right\rangle $ such as the one demonstrated in Ref. \cite{Kwiat}%
, the latter state can produced by straightforward polarization rotations %
\cite{Kwiat}. \ A linear optics implementation of Hardy's paradox (but not
the weak measurements) was proposed by Hardy in 1992 \cite{Hardyproposal}
and was recently demonstrated in Ref. \cite{Bouwmeester} subsequent to the
publication of Ref.\cite{Ahnert}.

\section{Two-particle Weak Measurement}

The aim of Ahnert and Payne was to suggest a feasible implementation of
Aharonov's weak measurements. \ In Aharanov's work, the single particle weak
measurements are simply identical to the IFM results. \ It is the
two-particle weak measurements that are most significant since they reveal a
consistent resolution of Hardy's paradox. \ In this section, we address the
two ways in which the paper proposes to do two-particle weak measurements. \ 

First, we review Ahnert and Payne's single-particle weak measurements. \
They base much of their apparatus and theory on an earlier paper \cite%
{Arrival}\ which employs the arrival time of the photon as a measurement
pointer similar to in Ref. \cite{Chiao}. \ They claim, their apparatus
measures the operator $A_{i}=\gamma \left| V_{i}\right\rangle \left\langle
V_{i}\right| +\varepsilon \left| H_{i}\right\rangle \left\langle
H_{i}\right| ,$ where $i=$ 1 or 2 indicates the particle number. \ In
contrast, Aharonov discussed the weak values of the polarization projectors $%
P_{Hi}=\left| H\right\rangle _{i}\left\langle H\right| _{i}$\ or $%
P_{Vi}=\left| V\right\rangle _{i}\left\langle V\right| _{i}$. \ The weak
value of $A_{i}$ is related to the weak values Aharonov discussed by $%
\left\langle A_{i}\right\rangle _{w}=\gamma \left\langle P_{Vi}\right\rangle
_{w}+\varepsilon \left\langle P_{Hi}\right\rangle _{w}$, where the subscript 
$w$\ indicates a weak value \cite{Aharanov}. \ With the identity\ $%
P_{Hi}+P_{Vi}=I$ one can derive the additional property of these weak values,%
\emph{\ } $\left\langle P_{Vi}\right\rangle _{w}+\left\langle
P_{Hi}\right\rangle _{w}=1$, which along with $\left\langle
A_{i}\right\rangle _{w}$, is sufficient to extract $\left\langle
P_{Vi}\right\rangle _{w}$ and $\left\langle P_{Hi}\right\rangle _{w}$. \ The
authors find that in Hardy's paradox $\left\langle A_{i}\right\rangle
_{w}=\varepsilon $, from which we can infer that $\left\langle
P_{Vi}\right\rangle _{w}=1$ and $\left\langle P_{Hi}\right\rangle _{w}=0.$ \
\ These results agree with the subsequent IFMs (as they must) and indicate
that, individually, each photon is in the inner arm (V polarization) of its
respective interferometer.

In their first method of two-particle weak measurement, the authors
represent the combined measurement of the two photons with an unconventional
vector operator. \ The problem with this type of operator is that in the
weak regime it measures quantities such as $\left\langle P_{V1}\right\rangle
_{w}$ and$\ \left\langle P_{V2}\right\rangle _{w}$ as opposed to $%
\left\langle P_{V1}P_{V2}\right\rangle _{w}$, one of the Aharonov's
two-particle weak values. \ The specific vector operator the authors propose
to measure is $A_{12}=(\gamma ,\gamma )\left| VV\right\rangle \left\langle
VV\right| +(\gamma ,\varepsilon )\left| HV\right\rangle \left\langle
HV\right| +(\varepsilon ,\gamma )\left| VH\right\rangle \left\langle
VH\right| +(\varepsilon ,\varepsilon )\left| HH\right\rangle \left\langle
HH\right| ,$ which, again, is not any one of Aharonov's four two-particle
projectors, such as $P_{VH}=\left| VH\right\rangle \left\langle VH\right| $,
but a combination of all four. \ The operator can be expanded as a vector 
\begin{eqnarray}
A_{12} &=&\left( 
\begin{array}{c}
\gamma \left| VV\right\rangle \left\langle VV\right| +\gamma \left|
HV\right\rangle \left\langle HV\right| +\varepsilon \left| VH\right\rangle
\left\langle VH\right| +\varepsilon \left| HH\right\rangle \left\langle
HH\right| , \\ 
\gamma \left| VV\right\rangle \left\langle VV\right| +\varepsilon \left|
HV\right\rangle \left\langle HV\right| +\gamma \left| VH\right\rangle
\left\langle VH\right| +\varepsilon \left| HH\right\rangle \left\langle
HH\right|%
\end{array}%
\right) \\
&=&\left( 
\begin{array}{c}
\left( \varepsilon \left| H_{2}\right\rangle \left\langle H_{2}\right|
+\gamma \left| V_{2}\right\rangle \left\langle V_{2}\right| \right) \left(
\left| H_{1}\right\rangle \left\langle H_{1}\right| +\left|
V_{1}\right\rangle \left\langle V_{1}\right| \right) , \\ 
\left( \varepsilon \left| H_{1}\right\rangle \left\langle H_{1}\right|
+\gamma \left| V_{1}\right\rangle \left\langle V_{1}\right| \right) \left(
\left| H_{2}\right\rangle \left\langle H_{2}\right| +\left|
V_{2}\right\rangle \left\langle V_{2}\right| \right)%
\end{array}%
\right) \\
&=&(\varepsilon \left| H_{2}\right\rangle \left\langle H_{2}\right| +\gamma
\left| V_{2}\right\rangle \left\langle V_{2}\right| ,\varepsilon \left|
H_{1}\right\rangle \left\langle H_{1}\right| +\gamma \left|
V_{1}\right\rangle \left\langle V_{1}\right| ) \\
&=&(A_{2},A_{1}).  \label{vector}
\end{eqnarray}%
On the surface, $A_{12}$\ appears to be a two-particle operator representing
the measurement of some joint-property of the two photons. In fact, Eq. \ref%
{vector} shows that $A_{12}$ can be re-expressed as a vector of two separate
single-particle operators, representing independent measurements of $A_{i}$
on each photon. \ Contrast this with the correct combination, $A_{1}\otimes $
$A_{2},$ that Aharonov employed in operators such as $\left| HV\right\rangle
\left\langle HV\right| .$ \ Therefore, their proposal, built upon the
operator $A_{12},$ is insufficient to measure any of the two-particle weak
values in Hardy's paradox. \ For example, the weak value for $\left\langle
A_{12}\right\rangle _{w}=(\left\langle A_{2}\right\rangle _{w},\left\langle
A_{1}\right\rangle _{w})=\left( \varepsilon ,\varepsilon \right) $ (Eq. 23
in \cite{Ahnert}) simply contains the weak values from the single-particle
weak measurements reviewed above$.$ \ If one incorrectly interprets this
result as the weak value for the location of the photon pair then one would
conclude that the photons were simultaneously in the inner arms (had
vertical polarizations). \ This is in direct contradiction with\emph{\ }%
Aharonov's prediction of 0 for the weak value $\left\langle
P_{V1}P_{V2}\right\rangle _{w}$, a prediction which necessarily must concur
with Hardy's paradox \cite{Aharanov} in that the photon pair is never found
in the inner arms simultaneously. \ 

Designing a linear optics experiment to implement the single-particle weak
measurements was never a major hurdle for Hardy's paradox. \ Measurement of
a single-particle weak value is straightforward when the pointer variable is
another degree of freedom in the quantum system. \ Examples include a
Stern-Gerlach device for measuring spin by coupling to the transverse
momentum of the particle, or the optical analogy with polarization \cite%
{Hulet}. \ Difficulties begin when one means to measure joint properties of
multiple particles. \ As with strong quantum measurements, weak values do
not obey a product rule, (i.e., $\left\langle AB\right\rangle _{w}\neq
\left\langle A\right\rangle _{w}\left\langle B\right\rangle _{w}$ \cite%
{Aharanov}) and so multiparticle weak values cannot be calculated from
single-particle ones. \ Instead, they must be measured. \ In the present
example, one has to weakly measure projectors such as $\left|
HH\right\rangle \left\langle HH\right| $. \ If one follows the original
approach to weak measurement, based on von Neumann system-pointer
interactions, nonlinear optical interactions at the two-photon level are
required to measure this projector. \ Recently, two of us devised a method
that avoids this obstacle \cite{Resch}\emph{\ }and is ideally suited for
linear optics. \ In that work, we show how one can indirectly extract joint
weak values by studying the correlations between two single-particle weak
measurements. \ This procedure has since been extended and simplified \cite%
{Lundeen}. \ In summary, a serious flaw in Ahnert and Payne's proposal is
that they do not outline any such indirect method for weakly measuring
two-particle joint properties while, at the same time, their apparatus is
incapable of directly measuring these properties.

Their second method of two-particle weak measurement consists of inserting
detectors in the apparatus before the post-selection. \ In the penultimate
paragraph of the paper, the authors assert that one can measure three of the
four two-particle weak values with this method. \ However, the detectors
would collapse each of the two photons to either H or V polarization\ and
strongly disturb the subsequent post-selection. \ This is a strong,
intrusive measurement and is exactly the situation that weak measurement is
designed to circumvent. \ Furthermore and not surprisingly, these
measurements give the strong measurement results, not the weak ones.

\section{Conclusion}

In conclusion, because of these flaws the paper does not outline a feasible
linear optics implementation of weak measurements in Hardy's paradox. \
However, the main obstacles to a linear optics implementation, state
preparation and two-particle weak measurements, have been described and
solved in other works in the literature.\ Consequently, there was already a
clear way to measure weak values in Hardy's paradox using only linear optics.

\begin{acknowledgments}
This work was supported by ARC Seibersdorf Research GmbH, the Austrian
Science Foundation (FWF), project number SFB 015 P06, NSERC, and the
European Commission, contract number IST-2001-38864 (RAMBOQ). We would like
to thank Terry Rudolph for helpful discussions.
\end{acknowledgments}

\end{document}